\documentclass{aims} 
\usepackage{amsmath}
\usepackage{paralist}
\usepackage[misc]{ifsym}

\usepackage[ruled,linesnumbered]{algorithm2e}
\usepackage{subfig}
\usepackage{epsfig} 
\usepackage{epstopdf} 
\usepackage[colorlinks=true]{hyperref}
\hypersetup{urlcolor=blue, citecolor=red}
\allowdisplaybreaks
\usepackage{cleveref}

\def\currentvolume{X}
 \def\currentissue{X}
  \def\currentyear{200X}
   \def\currentmonth{XX}
    \def\ppages{X-XX}
     \def\DOI{10.3934/xx.xxxxxxx}

\theoremstyle{definition}

\newcommand{\R}{\mathbb{R}}

\newcommand{\rev}[1]{{#1}}

\title[Reconstruction and segmentation from sequential measurements]{Reconstruction and Segmentation from sparse sequential X-Ray Measurements of Wood Logs}

\author[Sebastian Springer, Aldo Glielmo, Angelina Senchukova, et al.]{}


\thanks{This work was supported by the Centre of Excellence of Inverse Modelling and Imaging (CoE), Academy of Finland,  decision numbers 336787 and 336796. AH was was supported by the Academy of Finland project 338408. SS was supported by the Academy of Finland, project number 334 817.}

\thanks{$^*$Corresponding author: Sebastian Springer}

\begin{document}
\maketitle

\centerline{\scshape
Sebastian Springer$^{{\href{sebastian.springer@sissa.it}{\textrm{\Letter}}}*1}$,
Aldo Glielmo$^{{\href{mailto:aldo.glielmo@bancaditalia.it}{\textrm{\Letter}}}1}$,
Angelina Senchukova $^{{\href{mailto:angelina.senchukova@lut.fi}{\textrm{\Letter}}}2}$,
Tomi Kauppi $^{{\href{mailto:tomi.kauppi@finnos.fi}{\textrm{\Letter}}}3}$,}
\centerline{\scshape
Jarkko Suuronen $^{{\href{mailto:jarkko.suuronen@lut.fi}{\textrm{\Letter}}}2}$,
Lassi Roininen $^{{\href{lassi.roininen@lut.fi}{\textrm{\Letter}}}2}$, 
Heikki Haario $^{{\href{mailto:heikki.haario@lut.fi}{\textrm{\Letter}}}2}$,
and Andreas Hauptmann $^{{\href{mailto:andreas.hauptmann@oulu.fi}{\textrm{\Letter}}}45}$}
\medskip

{\footnotesize
 \centerline{$^1$International School for Advanced Studies (SISSA), Trieste, IT}

} 

\medskip

{\footnotesize
 \centerline{$^2$Computational Engineering Department, Lappeenranta-Lahti University of Technology, Lappeenranta, FI}
}

\medskip

{\footnotesize
 \centerline{$^3$Finnos Oy, Lappeenranta, FI}
   }

\medskip

{\footnotesize
 \centerline{$^4$Research unit of Mathematical Sciences, University of Oulu, Oulu, FI}
   }

\medskip

{\footnotesize
 \centerline{$^5$Department of Computer Science, University College London, London, UK}
   }

\bigskip

 \centerline{(Communicated by Handling Editor)}


\begin{abstract}
In industrial applications, it is common to scan objects on a moving conveyor belt. If slice-wise 2D computed tomography (CT)  measurements of the moving object are obtained we call it a sequential scanning geometry. In this case, each slice on its own does not carry sufficient information to reconstruct a useful tomographic image. Thus, here we propose the use of a Dimension reduced Kalman Filter to accumulate information between slices and allow for sufficiently accurate reconstructions for further assessment of the object. Additionally, we propose to use an unsupervised clustering approach known as Density Peak Advanced, to perform a segmentation and spot density anomalies in the internal structure of the reconstructed objects. 
We evaluate the method in a proof of concept study for the application of wood log scanning for the industrial sawing process, where the goal is to spot anomalies within the wood log to allow for optimal sawing patterns. Reconstruction and segmentation quality \rev{are} evaluated from experimental measurement data for various scenarios of severely undersampled X-measurements. Results show clearly that an improvement \rev{in} reconstruction quality can be obtained by employing the Dimension reduced Kalman Filter allowing to robustly obtain the segmented logs.
\end{abstract}



\section{Introduction}
Sequential scanning of dynamic processes or moving objects is a common practice in industrial applications of X-ray tomography \cite{de2014industrial, schoeman2016x}.
Examples are provided by an object moving along a conveyor belt and scanned by a stationary X-ray scanner, or by a dynamic process that is monitored in a scanning system. A common limitation in such configurations is that a full-scale rotational computed tomography (CT)  system \rev{\cite{ursella2018fast} may not \rev{be} available due to cost limitations}, but low-cost continual scanning modes are preferred, potentially with multiple sensor-detector pairs and slower rotational speed, \rev{if any \cite{schut2022top}.} In the case of a moving object on a conveyor belt, this leads to a sequential scanning geometry, in which only \rev{a} few measurements per two-dimensional slice can be obtained. The limited information may not be sufficient to reconstruct a high-quality tomographic image. Furthermore, if one aims to obtain a full three-dimensional reconstruction, then information between slices needs to be combined to provide sufficient angular information on the scanned object. 

In this study\rev{,} we aim to obtain an accurate two-dimensional tomographic image from only \rev{a} few measurement directions per slice by combining information between the slices. In particular, we evaluate the suitability of a Kalman filtering approach to accumulate measured data along the third dimension, e.g., the direction of movement along the conveyor belt. 
Usually, Kalman filtering is used to propagate information forward in time. Here, we interpret a sequential scanning of a three-dimensional object as an evolution of 2D slices along the third dimension, where the direction of movement serves as the temporal domain or simply a third dimension. As such, we consider this as a 2D+1 imaging scenario, especially applicable if structures change gradually along the third dimension. \rev{Many organic materials do so, and this is particularly true for the growth of trees. This} gives a natural analogy to a smooth temporal change in a Kalman model. 
%
Additionally, for most industrial applications, the crucial goal is not the tomographic reconstruction per se, but rather the identification \rev{of} relevant features, such as the location of foreign objects or inclusions.

In our study, we perform this identification step via a novel image segmentation scheme, obtained by appropriately modifying a recently published unsupervised clustering algorithm named Density Peaks Advanced (DPA) \cite{DERRICO2021476}. 

We evaluate the practical applicability of our approach for a wood log scanning scenario, where the logs move along a conveyor belt, and only \rev{a} few angle X-ray measurements can be taken for each sequentially scanned 2D slice. The aim is to obtain a sufficiently accurate reconstruction with as few angles as possible, by accumulating the information along the third dimension of the log. 
The secondary task is to accurately and reliably identify knots in the log to optimize a subsequent sawing process. Naturally, the tomographic image can be considered sufficient if all knots are identified correctly in this secondary segmentation task  \cite{Johansson:2013aa,Longuetaud:2012aa}. 
\rev{We note, that in contrast to the studies \cite{krahenbuhl2014knot,Longuetaud:2012aa} where \textit{a-posteriori} segmentation is performed on a dense angle full-log scan, we consider here a sparse sequential imaging scenario, where only a few slices at a time are available.}
\rev{Conceptually, the reconstruction approach considered here is} directly related to dynamic tomographic problems, where an object is assumed to evolve in time. Consequentially, the majority of related methods consider dynamic imaging problems that evolve over time and use \rev{the} information on the object's motion in the reconstruction task. Here we distinguish between motion compensation \cite{hahn2014efficient,hahn2021motion}, \rev{which} aims to reconstruct a reference state along with motion information, and dynamic reconstructions that provide a reconstruction of the target in the 2D+1 space-time cube \cite{burger2017variational,niemi2015dynamic}. We refer to \cite{hauptmann2021image} for a recent overview. Naturally, only methods of the latter, full dynamic type are relevant here. We provide an overview of related approaches in \cref{sec:tomoRec}. 

Furthermore, most reconstruction tasks are closely connected to a subsequent processing task. For instance, in medical imaging the CT image might be used for identification and quantification of cancer tissue \cite{tanoue2015lung}. Thus, in recent years some researchers have also increasingly considered to combine both tasks in a joint framework \cite{adler2021task,boink2019partially,arridge2021joint}. Nevertheless, we will concentrate here on a separated approach, but keep the segmentation quality in mind as evaluation criterion of reconstruction quality rather than  quantitative reconstruction errors.

This paper is organized as follows. In \cref{sec:tomoRec} we discuss the tomographic reconstruction task as well as the sequential setting and its relation to dynamic imaging. In \cref{sec:dimred_KF} we present our reconstruction framework based on a dimension reduced Kalman filter. Next we discuss the  use of the DPA method for the segmentation task in \cref{sec:segmentation}.  We present the experimental setup and results for different scanning geometries in \cref{sec:results}, followed by a discussion in \cref{sec:discussion}. Finally, we conclude in \cref{sec:conclusion}.

\section{Tomographic image reconstruction}\label{sec:tomoRec}
X-ray computed tomography represents a powerful non-destructive imaging technique that allows \rev{the assessment of objects} for quality control. X-rays are emitted and pass through the object of interest, the intensity of the X-ray beam changes governed by the Beer-Lambert law and the resulting intensity is recorded by the X-ray sensor. The obtained projections can be used to form a tomographic image in 2 or 3 dimensions, depending on the detection geometry. 

The main challenge connected to the tomographic reconstruction problem is related to its ill-posedness, as small changes in the data, usually caused by unavoidable measurement noise, can have a large impact on the reconstruction quality \cite{Kak2001,mueller2012linear}. The problem of ill-posedness is exacerbated when only sparse data is available. An additional challenge is to maintain computational feasibility for potential online reconstruction scenarios as both data and image can be high-dimensional.

In the following, we provide a mathematical formalization of the tomographic reconstruction problem. Here we consider the linearized problem, after log-transforming the recorded intensities. The linear forward mapping from the tomographic image $x\in X$ to the X-ray measurements $y\in Y$ then consists of integrating over straight lines \rev{travelling} through the target, where the geometry depends on the measurement system. 
Here we will consider a fan-beam geometry, that will be further explained in \cref{sec:results}. This measurement process can then be represented by the linear forward operator $A:X\to Y$, where $X$ and $Y$ are Hilbert spaces. The measurement model is then given by
\begin{equation}\label{eqn:measModel}
y = A x + \eta    
\end{equation}
where the $y\in Y$ is the measured sinogram that consists of X-ray projections for several measurement angles, $x\in X$ is the tomographic image to be reconstructed, and $\eta \in Y$ denotes measurement noise.

The corresponding inverse problem consists in reconstructing $x$ from the available measured sinogram $y$ under knowledge of the scanning geometry described by $A$. This needs to be done in a robust manner to prevent the noise $\eta$ \rev{from} corrupting the tomographic image. The most common reconstruction is given by filtered backprojection (FBP), which consists of a frequency filter of the measured data, followed by the backprojection operation given the adjoint $A^*:Y\to X$.
Unfortunately, the FBP assumes dense angular sampling and does not provide sufficient reconstruction results under sparse scanning geometries \cite{mueller2012linear}.

Alternative approaches are given by a variational formulation of the reconstruction problem \cite{scherzer2009variational}. That is, we aim to find the reconstruction $x_{\mathrm{rec}}$ as the minimizer of a cost functional. In the simplest quadratic case, the reconstruction can be obtained as
\begin{equation}\label{eqn:TikFunc}
    x_{\mathrm{rec}}=\arg\min_{x\in X} \frac{1}{2}\|Ax-y\|^2_2 + \alpha \|x\|_2^2,
\end{equation}
where $\alpha$ balances between the two terms. The first term measures the goodness-of-fit, or data-discrepancy, whereas the second term is the so-called regularizer ensuring convergent and robust reconstructions. A major advantage of the formulation in \eqref{eqn:TikFunc}, is that the solution can be represented in a closed form as
\begin{equation}\label{eqn:TikClosedForm}
    x_{\mathrm{rec}}=\left(A^T A + \alpha I \right)^{-1}A^* y,
\end{equation}
where $I$ is the identity. We note that other regularizers can be considered in \eqref{eqn:TikFunc}, \rev{which} encode different prior information on the expected targets. For instance, piece-wise constant reconstructions via total variation \cite{chambolle2011first,rudin1992nonlinear}, wavelet sparsity \cite{daubechies2004iterative}, or even learned regularizer \cite{lunz2018adversarial}.

\subsection{Sequential tomography}

It is commonly known, and easily verified by reconstructions of logs with high resolution,  that logs of wood are objects whose internal structure varies slowly along the vertical direction. 
Two consecutive sections of a log are almost always similar to each other if the distance between them is in the order of 2-5 \rev{millimetres}. 
This observation makes it obvious that the quality of the reconstructions obtained from sparse angle data could be improved by using the information of one or more neighboring slices.

Examples of approaches that fit this scenario are reconstruction algorithms for evolving targets. A classic way to propagate information forward is given by filtering approaches like Kalman Filters (KF), which are designed for estimating the states of time series models. Alternatively, one could also consider data-driven methods such as recurrent neural networks, if large amounts of high quality data are available. 
Here we will focus on the former as the availability of abundant high quality measurements is not always freely available.  We mimic the time series type of data by considering the wood log as a  sequence of 2D  slices that evolve as the log passes \rev{through} the scanner.


That means we do not aim at reconstructing one image or volume at a time as formulated in \cref{eqn:measModel}, but rather a series of reconstructions $x_k$ for $k=0,1,2,3 ,...$. Here, each $x_k$ corresponds to a 2D image and a 3D volume is obtained by a stack of 2D reconstructions. The essential task of the dynamic reconstruction problem is now to use the relation between slices $x_k$ efficiently to compensate for possible highly sparse measurements for each $k$ separately. Naturally, in order to accumulate the information along the third dimension, the measurement geometry, i.e., the angular sampling, needs to change. 
That means the forward operator changes for each slice. The sequential measurement model is then given by
\begin{equation}\label{eqn:measModel_k}
y_k = A_k x_k + \eta_k.
\end{equation}
At each slice $x_k$ we take a different set of measurements \rev{modelled} by $A_k$ and obtain the corresponding measurement $y_k$ under noise $\eta_k$. 


\section{Dimension Reduced Kalman-filter}\label{sec:dimred_KF}

There exist a vast amount of filtering approaches. For reasons of computational efficiency we select to build our model on the previously introduced Dimension Reduced Kalman Filtering (DrKF)  method \cite{Janne19,Antti16}. In DrKF,  the X-ray reconstructions are parameterized by a low-dimensional basis, that reduces the dimensionality of the state space, and consequently decreases the computational cost. In the following we consider the finite dimensional case, that is $x\in \R^N$, $y\in\R^M$ and the forward operator is given as matrix $A:\R^N\to\R^M$. 
Let us now give a brief summary of the method considered, for further details see \cite{Janne19,Antti16}. That is, we consider the following pair:
\begin{align}
x_k &=M_k x_{k-1} + \epsilon_k \\
y_k &= A_k x_k + \eta_k
\end{align}
 where $x_k$ is the $k$-th section of our scanned log; $M_k=I$ is our forecast operator, moving from one slice to the next one; $\epsilon_k$ is a zero mean Gaussian model error with covariance matrix \rev{$Q_k\in \R^N$}; $y_k$ is the sinogram data; the observation model $A_k$ is the corresponding discretized matrix from \cref{eqn:measModel_k} and $\eta_k$ is a zero mean Gaussian observation error with covariance matrix \rev{$R_k\in \R^M$}.

\rev{Finding the model error covariance matrix is a crucial step in ensuring the accuracy of the filtering process. Through careful trial and error, we have found a $Q_k$ that balances between giving sufficient weight to new measurement information and not forgetting the valuable information from the previous reconstructions. }

For $k = 0$ we have $x_0$ as in \cref{eqn:TikClosedForm} with $A$ substituted by the projected form $A_0 P_r$, which we will explain next. 
The idea of dimension reduction is to constrain the problem into a subspace that \rev{contains} most of the variability allowed by the prior. The projection matrix \rev{$P_r\in R^N \times R^r$} used for prior-based dimension reduction is obtained from the prior covariance matrix $\Sigma$ using the singular value decomposition. In this notation, $r$ stands for the $r$ leading singular values used.
\rev{The covariance matrix $\Sigma$ in this case is obtained by using standard Gaussian covariance function: 
$\Sigma_{i,j} = \sigma^2 \exp{ - \frac{d(x_i,x_j)^2}{2l^2}},$ where $\sigma^2$ is the variance parameter, $l$ is the correlation length and $d(x_i, x_j )$ is the Euclidean distance between pixels $x_i$ and $x_j$. For practical purposes we select $\sigma = 0.1$ and $l = 1.5$.}
\rev{It is possible to use a different prior covariance and achieve comparable outcomes. However, we chose this specific prior covariance because it was shown to produced visually satisfactory reconstructions, for further details, we refer to the experimental section of \cite{Janne19}.}
One can compute \rev{the SVD of the covariance matrix as} $\Sigma=U S U^T$ where U is a unitary matrix containing the singular vectors and S is a diagonal matrix containing the singular values.
The projection matrix $P_r$ can be obtained as:
\begin{equation}
P_r = U_r S_r^{1/2}    
\end{equation}

Let us now parametrize our images using the projection matrix $P_r$ as
 \begin{equation}
     x_k=x_k^p + P_r \alpha_k
 \end{equation}
and thus obtain the following prediction step
\begin{align}
    x_k^p &= M_k ( x_{k-1}^p + P_r \alpha_{k-1}^{est} ) \label{eqn:x_update}
    \\
    C_k^p &= (M_k P_r) (\phi_{k-1}^{est}) (M_k P_r)^T +Q_k
\end{align}
where $x_k^p$ is the prediction mean and $C_k^p$ is the prediction covariance matrix.
The time consuming update step is performed in the lower dimensional space, 
\begin{align}
    \alpha_{k}^{est} &=\phi_{k}^{est} (A_k P_r)^T R_k^{-1} (y_k - A_k x_k^p)\label{eqn:alpha_k} \\
    \phi_{k}^{est} &=  (  (A_k P_r)^T R_k^{-1} (A_k P_r) + P_r^{T} (C_k^{p})^{-1} P_r    + \xi I )^{-1} \label{eqn:phi_k}
\end{align}
We note that the above equations are conceptually similar to \cref{eqn:TikClosedForm}, to produce the update for $x_k^p$ in \cref{eqn:x_update}.
To further reduce the computations one can observe that $C_k^p = B_kB_k^T + Q_k$, $B_k= M_k P_r V_k$, $\phi_{k-1}^{est}=V_k V_k^T$ and therefore obtain:
\begin{equation}
(C_k^{p})^{-1} P_r = Q_k^{-1} P_r  - Q_k^{-1} B_k (B_k^T Q_k^{-1} B_k +I)^{-1} B_k^T Q_k^{-1} P_r.\label{eqn:C_k}
\end{equation}


Additionally, in the computation of $\phi_{k}^{est}$ we introduced a regularizer here before taking the inverse to avoid getting a singular matrix, this showed to be necessary for the highly sparse imaging scenarios under consideration here. The regularizer is given by the identity matrix multiplied by a constant that grows linearly with slope $\xi=0.1$ as the number of angles \rev{increases}. 
\rev{We note, that due to the nature of the Kalman filter, the first 5-10 slices are needed to accumulate information, after that we can expect good reconstruction performance. A full reconstruction pipeline is given Algorithm \ref{alg:1}.}


\begin{algorithm}
\SetAlgoLined
\caption{Pipeline of the reconstruction workflow }
\label{alg:1}
         \SetKwInOut{Input}{input}
        \SetKwInOut{Output}{output}
        \Input{ Sinograms $\{y_k\}_{k=0}^N$, system matrix $A$, number of slices $N$}
        \Output{  Reconstructed $\{x_k\}_{k=0}^N$}
    \SetKwBlock{Beginn}{beginn}{ende}
\Begin{
            Select reduced dimension $r << N$;\\
            Select prior covariance matrix $\Sigma$, e..g: \\
            $\Sigma_{i,j} = \sigma^2 \exp{ - \frac{d(x_i,x_j)^2}{2l^2}}$, with $\sigma^2$ the variance parameter, $l$ the correlation length and $d(x_i, x_j )$, and the Euclidean distance between pixels $x_i$ and $x_j$;\\
            Construct Projection matrix $P_r = U_r S_r^{1/2}$, by the SVD on the covariance matrix $\Sigma=U S U^T$ ;\\
            Reconstruct the first slice $x_0$ in the full space;\\
            N $\leftarrow$ number of slices;\\
        \For{$k =1$ to $N$}{
               Compute the prior mean $x_k^p$ by \eqref{eqn:x_update};\\
               Perform the decomposition $\phi_{k-1}^{est}=V_k V_k^T$;\\
               Compute the matrix $B_k= M_k P_r V_k$;\\
               Compute $(C_k^{p})^{-1} P_r$ by \eqref{eqn:C_k};\\
               Compute $\alpha_k^{est}$ and $\phi_{k}^{est}$ by eqns. \eqref{eqn:alpha_k} and \eqref{eqn:phi_k};\\
        }

    \Return  $\{x_k\}_{k=0}^N$
    }
\end{algorithm}


\section{Segmentation}\label{sec:segmentation}

In the literature, one can find many articles \cite{ Andreu:2003aa, Baumgartner:2010aa, Hagman:2003aa,Hagman:1995aa, Nordmark:2002aa, Todoroki:2010aa} and reviews \cite{Gergel:2019aa,Longuetaud:2012aa}  describing various kinds of approaches that can be used to locate specific structures like knots, cracks, resin pockets and alien objects inside logs from high quality reconstructions obtained using low speed CT scanning. 
Most of these approaches have a knot detection rate exceeding $90\%$ with about $1\%$ falsely detected knots.

However, in contrast to standard \rev{measurement} settings, this work focuses on severely sparse angular measurements and levels of perturbations much different \rev{from} those of medical CT scanners. To deal with this scenario, we propose to use an unsupervised segmentation approach particularly robust against noise.

\subsection{Segmentation by Density Peaks Advanced (DPA)}\label{sec:segmentation_dpa}

We consider a fully unsupervised clustering approach, the Density Peaks Advanced (DPA) proposed in \cite{DERRICO2021476}.

\rev{The method can be used to segment the reconstructed log either slice by slice (in 2D) or directly in 3D, and in the experiments presented in \cref{sec:results} all segmentations have been done directly on blocks of $11$ consecutive slices (3D).}\\

\rev{The main intuition behind the DPA procedure is to identify statistically significant clusters of points as as statistically significant `density peaks', i.e., groups of points that have high density and that and are separated by regions of low density. }

\rev{Specifically, DPA first estimates the local density around each point using the Point Adaptive k-nearest neighbour (PAk) \cite{Rodriguez18} estimator, and then identifies statistically significant clusters using the procedure of the Density Peaks (DP) algorithm \cite{Rodriguez1492} modified with three heuristics}

\rev{For our application, we do \emph{not} need to use the PAk density estimator, since X-ray intensities can be directly taken as estimates of the material densities in the same location, and hence modify the original DPA scheme as follows. For further details on the DPA scheme see \cite{DERRICO2021476}.}

\rev{Given the intensity values in each pixel of the tomographic image, DPA needs estimates of a log-density $\log(\rho_i)$ around each point $i$, the error on the log-density $\zeta_i$ and the number of neighbours $\hat{k}$ among which the density is assumed to vary only slightly.}

\rev{In our case, we assign to the variable $\rho$ the approximation of the densities obtained by our reconstructions $x$ and we estimate the noise level $\zeta_i$ from the background noise surrounding the reconstructed log (with mean $0.033$ and standard deviation $0.003$), while we fix the neighbourhood size to 200 as we empirically found that it is a good compromise between classifying correctly small but still relevant higher-density areas and disregarding density fluctuations.}. 

The next step in the procedure is to find automatically the density peaks (or `clusters')  of the image.
This can be done by observing that density peaks are surrounded by neighbours with lower local density but at relatively large \rev{distances} from points with higher local density.

To make our estimates more robust against noise, cluster \rev{centres} are defined as points that \rev{maximize} locally, and in a statistically significant manner, the error-scaled log-density $g_i=\log(\rho_i) - \zeta_i$. This is done as follows.

We start by finding a preliminary cluster assignment using \\
\emph{Heuristic 1}: A point $i$ is a local density \rev{centre} if all its $\hat{k}$ nearest neighbours have a value of $g$ lower \rev{than} $g_i$ and it does not belong to the neighbourhood of any other point with higher~$g$. 
After all the \rev{centres} have been found, all the pints \rev{(in our specific case voxels of the consecutive reconstructed slices)} are assigned to the same cluster of the nearest point with \rev{a} higher cost function value.

The next task is to find the points on the border between two clusters using \\
\emph{Heuristic 2}: A point $i$ of the cluster $c$ is defined to be on the boundary between the two clusters  $c$ and $c'$ if its closest point $j$ of the cluster $c'$ lies within the distance $r_{\hat{k}}$ given by its neighbourhood $\hat{k}$. Moreover, it must be the closest to $j$ among those belonging to $c$.
The saddle point between two clusters $c$ and $c'$ is defined as the point with \rev{a} higher value of $g$ between those on the boundary between the two clusters.

The log-density of the saddle point and its error are then defined respectively as $\log(\rho_{cc'})$ and $\zeta_{cc'}$. Based on these last two quantities it is possible to write a criterion for distinguishing genuine density peaks from false peaks generated from measurement errors using\\ 

\emph{Heuristic 3}: A cluster $c$ is considered to be the result of a density fluctuation if all its values have comparable density values to the one on the border. In particular,   cluster $c$ is merged with the cluster $c'$ if
$$(\log(\rho_c) - \log(\rho_{c c'}) ) < Z ( \zeta_c + \zeta_{c c'} ) $$
where $\rho_c$ is the density of the center of the cluster $c$, and the constant $Z$ fixes the level of statistical confidence. Heuristic 3 is checked for all the clusters $c$ and $c'$ in decreasing order of $\log(\rho_{cc'})$. 

The segmentation map obtained contains statistically significant clusters (each labelled with increasing positive numbers), which will be characterised by grid points having \rev{a higher density of image intensities}. All the \rev{others} will be assigned to the so called `halo' cluster (labelled as -1).
As our goal was only to spot density anomalies within the logs we will merge together all the statistically significant clusters by assigning all of them to the same cluster.

\subsection{Implementation}

The described segmentation method is implemented in Python using the DADApy toolbox \cite{DADApy}, while the DrKF reconstruction part is done in Python by using standard libraries.
As stated previously the main difference with respect to the examples given with the DADApy toolbox is that we start from the approximation of the densities obtained by our reconstructions, we use the statistic of the background noise to assign a noise value to each grid point and we fix the neighbourhood size to separate genuine knots from density fluctuations, while in general the toolbox would estimate all the values from point clouds of data.

\section{Wood log reconstruction and segmentation}\label{sec:results}

In the sawing process, the logs are cut into boards. 
Summarizing in one sentence, we can say that optimizing this process consists  \rev{of} producing the highest possible quality boards from each log, minimizing the waste of time and resources. 
In this context, quality is directly correlated with economical value and therefore it is important to estimate the potential of each wood log in terms of boards that can be obtained from it. To do so, the following must be considered: one must know the internal structure of each log, locate all the elements that can influence the product quality (e.g. cracks, knots, foreign objects, etc.), and finally optimize the cutting parameters such as angle and dimensions to maximize the potential value of the extracted boards. 

The first step is therefore getting accurate CT images from sparse sequentially obtained projection data. The reason why there is a need for methods that can work well with \rev{a} few angles is mainly economical, as every extra X-ray source added has a significant impact on the cost of the measurement device and consequently also an overhead cost added to the sawing process.

In this section, we will study, within the proposed filtering framework, how reconstruction quality decays as the number of sources decreases for different acquisition schemes. The comparison will be done on the reconstruction and the corresponding segmentation, where our reference will be the reconstruction obtained with standard methods and dense X-ray measurements.

\subsection{Acquisition geometry and data calibration}

\begin{figure}[ht]
    \centering
    \includegraphics[width=0.6\linewidth]{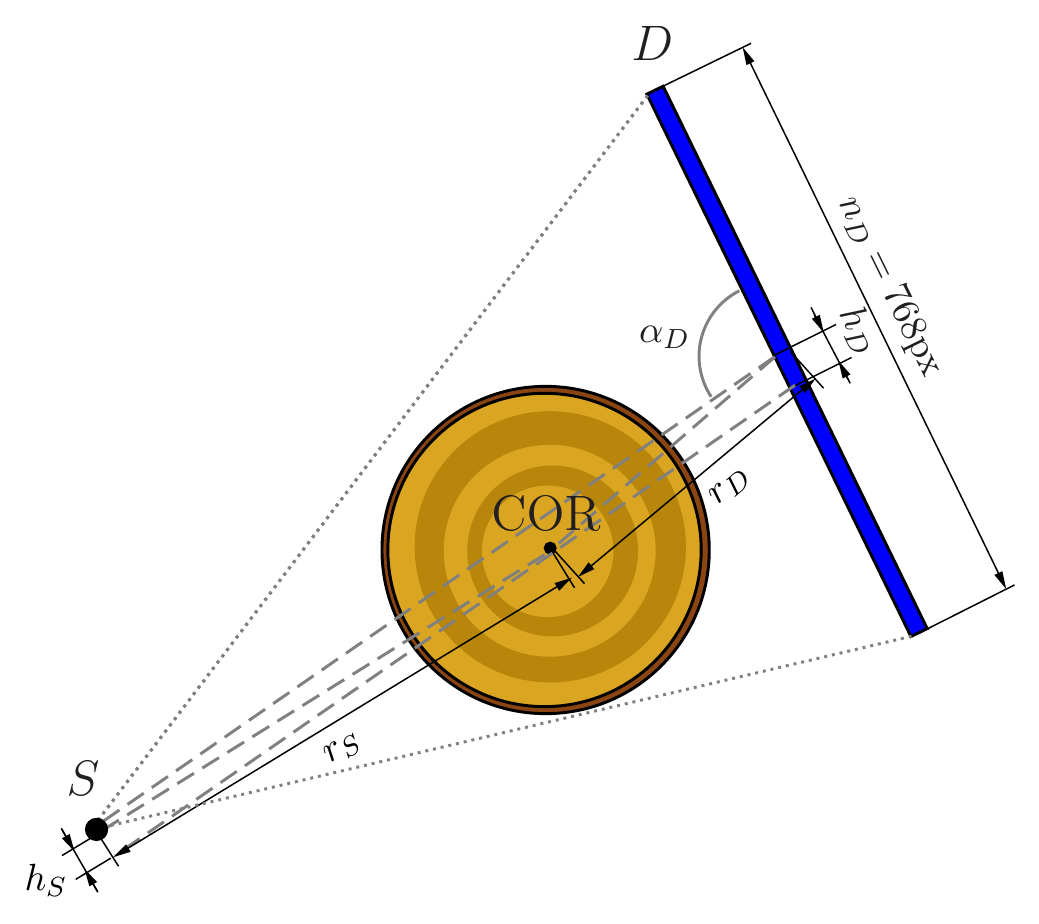}
    \caption{    A schematic drawing of the fan-beam acquisition geometry.
    Here $S$ represents one of the X-ray sources while $D = 1154.2 $ is the plate on which detectors were installed. The parameter $r_{S} = 859.46 $ is the distance between the X-ray source and \rev{centre} of rotation, $r_{D} = 705.37$ is the distance between the  \rev{centre} of rotation and the detector, $h_{S} = 232.86$ is the source shift, $h_{D} = -24.65  $ is the detector shift, $\alpha_{D} = 0.16$ is the detector tilt, and $n_{D} = 768$ is the number of detector elements. }\label{fig:x-ray_setup}
\end{figure}

The data used in this work has been produced by the Finnish company Finnos~Oy, specialized in solutions for \rev{the} sawing industry \cite{FinnosOy}. 
A schematic representation of the tomographic set-up used to produce the data we used can be found in \cref{fig:x-ray_setup}. The X-ray scanner, fixed around the conveyor belt along which the log is moved and rotated, is represented by a source emitting the fan-shaped X-ray beams and a corresponding flat line detector which consists of $768$ pixels that \rev{record} the beam intensities.

\begin{figure}[!ht]
    \centering

        \subfloat[Calibration phantom 1]{\includegraphics[width=0.32\columnwidth]{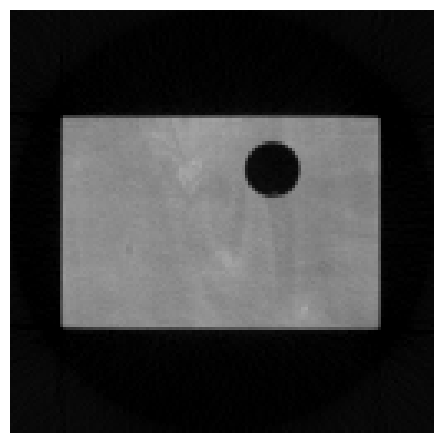}}
        \hspace{0.1cm}
        \subfloat[Calibration Phantom 2]{\includegraphics[width=0.32\columnwidth]{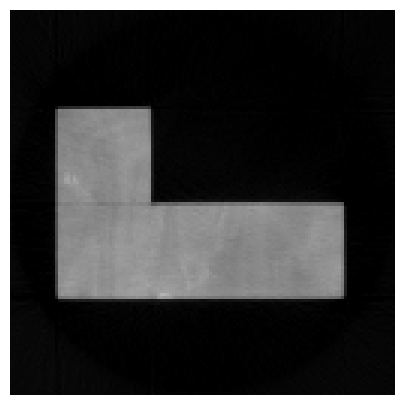}}
        \caption{Wooden phantoms used to perform the system matrix calibration }
    \label{fig:calibration phantoms}
\end{figure}

We define as $\Delta$ the angular difference between consecutive sources in the acquisition geometry.
To obtain the exact geometric values shown in \cref{fig:x-ray_setup} we used a calibration procedure using two wooden phantoms \cref{fig:calibration phantoms}, which were glued on both ends of the scanned trunk to calibrate the parameters of our virtual geometry implemented in python with ODL \cite{Adler_odl} using ASTRA \cite{van2016fast}.

As \rev{a} reference object in this work, we selected a spruce trunk. We scanned the trunk with dense angular measurements every 2mm in the vertical directions so as to tune the respective system matrixes and obtain our reference reconstructions using FBP.
We then sub-sampled the respective sinograms to use them as data for our approach.  

We note that despite all the precautions one can introduce to maintain the consistency of the data, a certain uncertainty remains in the parameters of the fan-beam geometry due to different sources of vibrations and mechanical displacements introduced while the log is moved on the conveyor belt. Thus advanced and robust approaches for reconstruction are indeed necessary.

\subsection{Reconstruction and Segmentation}

In this section, we will examine how the quality of the reconstruction and subsequent segmentation decreases as the number of sources decreases for different rotational schemes. We refer here to rotation regardless of whether it concerns the log or the measurement system, as it can be implemented either way in practice.
First, we will show how rotation plays a fundamental role in the quality of the reconstruction. In the absence of rotation, we will see how there is practically no improvement in the quality of the reconstruction when using the DrKF.
On the other hand, even if rotation occurs in very small increments the situation improves, but the internal nodes are still not adequately reconstructed.
The situation is different if the angle between measurements $\Delta $ is sufficiently large or even randomized. In that case, it is possible to obtain a quality of reconstruction sufficient to obtain a decent segmentation even with only 3 angles per slice.

The image resolution adopted in the following experiments was $128^2$ while the lower dimensional space of the filtering approach was chosen as $r=3000$, which is about $\sim 18\%$ of the original problem size. The trade-off between reconstruction quality and computational speed can be modified by increasing or decreasing $r$ according to the specific needs.

\subsubsection{Reconstructions with fixed acquisition geometry }

\begin{figure}[!ht]
    \centering
        \subfloat[FBP reference: slice 217 ]{\includegraphics[width=0.35\columnwidth]{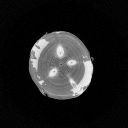}}
        \subfloat[DrKF reconstruction 217 ]{\includegraphics[width=0.35\columnwidth]{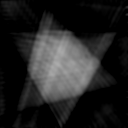}}
        \hspace{0.01cm}
        \subfloat[FBP reference: slice 232 ]{\includegraphics[width=0.35\columnwidth]{figures/reco_sl232_fbp_360_sources}}
        \subfloat[DrKF reconstruction 232 ]{\includegraphics[width=0.35\columnwidth]{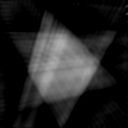}}
    \caption{ (a) and (c) are the reference figures obtained using FBP and dense sampling with 360 measurement angles while (b) and (d) are the respective reconstructions obtained using DrKF and a fixed system of 3 X-ray source-receiver pairs. }           
    \label{fig:DrKF_no_Rotation}
\end{figure}

In the first result presented we will use as reference 15 consecutive slices taken from one scanned log. We start our computation
from a fixed first  slice number 217 and proceed with the DrKF formulas by
keeping the system matrix fixed, namely in absence of rotation. The first and last of these 15 reconstructions are presented in  \cref{fig:DrKF_no_Rotation} in comparison the the FBP from full angular data. As can be seen, there has not been any significant accumulation of information after 15 iterations of the filtering formulas. Moreover, the quality of the reconstructions does not allow for the segmentation method used to locate the knots inside the wood log.

\subsubsection{Reconstructions with rotating acquisition geometry}

\begin{figure}[!ht]
    \centering

     \subfloat[Reference A ]{\label{fig:DrKF_REFERENCES_A} \includegraphics[width=0.32\columnwidth]{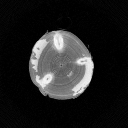}}
       \vspace{0.01cm}
      \subfloat[Reference B ]{\label{fig:DrKF_REFERENCES_B} \includegraphics[width=0.32\columnwidth]{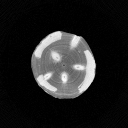}}
      \caption{ Two reference slices used to compare the reconstruction/segmentation quality.  }
\end{figure}

From now on we will consider only cases with rotating acquisition geometry. As \rev{a} reference, we will have two sections of 11 slices \rev{centred} respectively at  \cref{fig:DrKF_REFERENCES_A} and  \cref{fig:DrKF_REFERENCES_B}. We will use a burn-in period for the Kalman filter of 50 slices before the reference blocks are computed to have consistent and comparable reconstruction quality between the different imaging geometries.

We start by considering the case in which 
two consecutive acquisition geometries have a random angular increment, and we present reconstructions with \rev{an} increasing odd number of sources from 1 to 15. Where the difference between consecutive sources $\Delta$ is uniformly chosen over 360 degrees. \rev{As an example for $5$ sources these will be located at $( 0, 60, 120, 180, 240 ) $ and an angular increment of $35$ degrees we will move these to $( 35, 95, 155, 215, 275 ) $ in the next slice while a further angular increment by $42$ degrees would result in $( 77, 137, 197, 257, 317)$ in the second one.
We, note that the randomness is only partial as we ensure covering the full angular range before the sampling scheme can return to the initial state.}

\begin{figure}[!ht]
    \centering
        \subfloat[DrKF 1 source ]{\includegraphics[width=0.32\columnwidth]{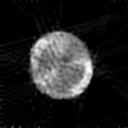}}
        \hspace{0.1cm}
        \subfloat[DrKF 3 sources ]{\includegraphics[width=0.32\columnwidth]{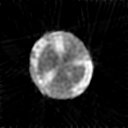}}
        \hspace{0.1cm}
        \subfloat[DrKF 5 sources ]{\includegraphics[width=0.32\columnwidth]{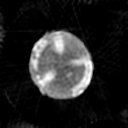}}
\hspace{0.1cm}
        \subfloat[DrKF 7 sources ]{\includegraphics[width=0.32\columnwidth]{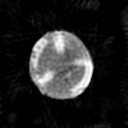}}
\hspace{0.1cm}
        \subfloat[DrKF 9 sources ]{\includegraphics[width=0.32\columnwidth]{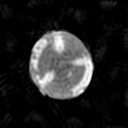}}
    \caption{DrKF reconstructions of the Reference A with acquisition geometry rotating with random speed.  }
    \label{fig:DrKF_RDrA}
\end{figure}

The visual inspection of the resulting reconstructions for Reference A presented in \cref{fig:DrKF_RDrA} indicates that the reconstruction quality is already good when using only 3 sources and does not increase significantly if one \rev{uses} more than 5. 
This is further confirmed \rev{by} the Peak signal-to-noise ratio (PSNR) \cite{rao2018transform} values given in \cref{fig:coeff_comparisons}.

\begin{figure}[!ht]
    \centering

        \subfloat[DrKF 1 source ]{\includegraphics[width=0.32\columnwidth]{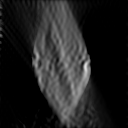}}
        \hspace{0.1cm}
        \subfloat[DrKF 3 sources ]{\includegraphics[width=0.32\columnwidth]{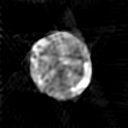}}
        \hspace{0.1cm}
        \subfloat[DrKF 5 sources ]{\includegraphics[width=0.32\columnwidth]{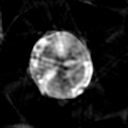}}
        \hspace{0.1cm}
        \subfloat[DrKF 7 sources ]{\includegraphics[width=0.32\columnwidth]{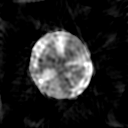}}
        \hspace{0.1cm}
        \subfloat[DrKF 9 sources ]{\includegraphics[width=0.32\columnwidth]{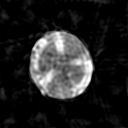}}
\caption{DrKF reconstructions of the Reference A with acquisition geometry rotating by 1 degree between consecutive slices. }
    \label{fig:DrKF_1DrA}
\end{figure}

Next follows the case in which the scanner is rotating at a constant slower speed so that every consecutive reconstruction is obtained with an acquisition geometry rotated by just 1 degree with respect to the previous one.
In this case, the reconstruction quality is clearly lower as can be seen from  \cref{fig:DrKF_1DrA} and it is further confirmed from the PSNR coefficients given in  \cref{fig:coeff_comparisons} as the values are all lower \rev{than} the 3 sources reconstruction given in  \cref{fig:DrKF_RDrA} .

\begin{figure}[!ht]
    \centering
        \subfloat[DrKF 1 source ]{\includegraphics[width=0.32\columnwidth]{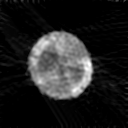}}
        \hspace{0.1cm}
        \subfloat[DrKF 3 sources ]{\includegraphics[width=0.32\columnwidth]{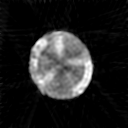}}
        \hspace{0.1cm}
        \subfloat[DrKF 5 sources ]{\includegraphics[width=0.32\columnwidth]{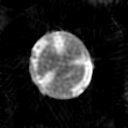}}
\hspace{0.1cm}
        \subfloat[DrKF 7 sources ]{\includegraphics[width=0.32\columnwidth]{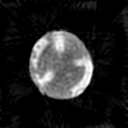}}
\hspace{0.1cm}
        \subfloat[DrKF 9 sources ]{\includegraphics[width=0.32\columnwidth]{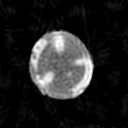}}
    \caption{DrKF reconstructions of the Reference A with acquisition geometry rotating by $\Delta/4$  degree between consecutive slices, where $\Delta$ is the angular difference between consecutive sources in the acquisition geometry.  }
    \label{fig:DrKF_4DrA}
\end{figure}

\begin{figure}[!ht]
    \centering
        \subfloat[DrKF 1 source ]{\includegraphics[width=0.32\columnwidth]{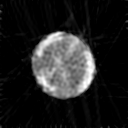}}
        \hspace{0.1cm}
        \subfloat[DrKF 3 sources ]{\includegraphics[width=0.32\columnwidth]{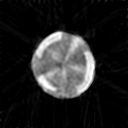}}
        \hspace{0.1cm}
        \subfloat[DrKF 5 sources ]{\includegraphics[width=0.32\columnwidth]{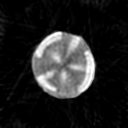}}
\hspace{0.1cm}
        \subfloat[DrKF 7 sources ]{\includegraphics[width=0.32\columnwidth]{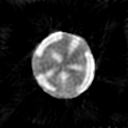}}
\hspace{0.1cm}
        \subfloat[DrKF 9 sources ]{\includegraphics[width=0.32\columnwidth]{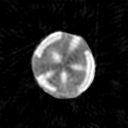}}
    \caption{DrKF reconstructions of the Reference B with acquisition geometry rotating by $\Delta/4$ degree between consecutive slices, where $\Delta$ is the angular difference between consecutive sources in the acquisition geometry.  }
    \label{fig:DrKF_4DrB}
\end{figure}

The last type of acquisition scheme presented \rev{consists} of a constant rotation with increment given by the closest entire number to 1/4 of the angular difference between two sources $\Delta$, which is not a divisor of $\Delta$. In this way, it takes 5 iteration steps (approximately 1 cm within the log) before returning to a set of angles that is close to the initial but rotated slightly\rev{, this sampling scheme is motivated by the golden angle sampling scheme in magnetic resonance imaging, that aims to sample the whole Fourier space radially.} For example for the 5 sources case described above the rotation angular increment is given by 16.
Both the visual inspection of the resulting reconstructions for Reference A presented in \cref{fig:DrKF_4DrA} and the PSNR coefficients of  \cref{fig:coeff_comparisons} indicate that this scheme produces similar results to those of the random angular rotation.
We performed the same test for Reference B and the reconstructions presented in  \cref{fig:DrKF_4DrB} \rev{lead} to similar conclusions.

\subsubsection{Segmentation comparison}

\begin{figure}[!ht]
\centering
\subfloat[Reference A]{\includegraphics[width=0.32\columnwidth]{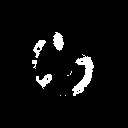}}
\hspace{0.1cm}
\subfloat[DrKF 1 source ]{\includegraphics[width=0.32\columnwidth]{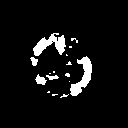}}
\hspace{0.1cm}
\subfloat[DrKF 3 sources ]{\includegraphics[width=0.32\columnwidth]{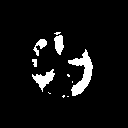}}
\hspace{0.1cm}
\subfloat[DrKF 5 sources ]{\includegraphics[width=0.32\columnwidth]{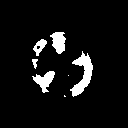}}
\hspace{0.1cm}
\subfloat[DrKF 7 sources ]{\includegraphics[width=0.32\columnwidth]{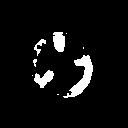}}
\hspace{0.1cm}
\subfloat[DrKF 9 sources ]{\includegraphics[width=0.32\columnwidth]{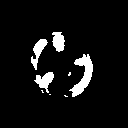}}
\hspace{0.1cm}
\subfloat[DrKF 11 sources ]{\includegraphics[width=0.32\columnwidth]{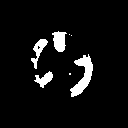}}
\hspace{0.1cm}
\subfloat[DrKF 13 sources ]{\includegraphics[width=0.32\columnwidth]{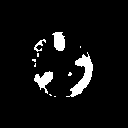}}
\hspace{0.1cm}
\subfloat[DrKF 15 sources ]{\includegraphics[width=0.32\columnwidth]{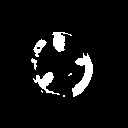}}
\caption{Segmentations of the Reference A and the reconstructions presented in  \cref{fig:DrKF_4DrA}.}
\label{fig:segm4}
\end{figure}

\begin{figure}[!ht]
    \centering
        \subfloat[Average Dice squared coefficient. ] {\includegraphics[width=0.45\columnwidth]{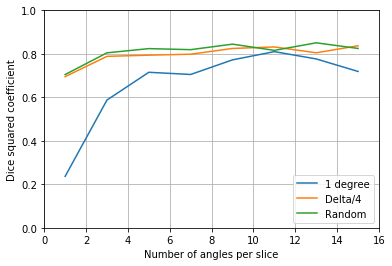}}
        \hspace{0.1cm}
        \subfloat[Average PSNR.]{\includegraphics[width=0.45\columnwidth]{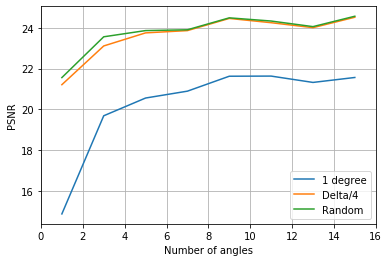}}

    \caption{Comparison of quantitative values for for different acquisition schemes. Values are obtained by averaging in the block of 11 slices centered at the reference in fig. \ref{fig:DrKF_REFERENCES_A}. (A) Dice squared coefficient, where the reference segmentation was obtained from the full angular FBP reconstruction.
    With $Z$ set to 3.4 for number of angles smaller \rev{than} 7 and 2.4 for the other cases. (B) Average PSNR.}
    \label{fig:coeff_comparisons}
\end{figure}

We then segmented all the reconstructions presented previously to assert the ability of the segmentation method to spot the location, size and direction of the knots within the wood logs. 
\Cref{fig:segm4} contains both the \rev{segmentations} of the reference reconstruction given in  \cref{fig:DrKF_REFERENCES_A} and those given in  \cref{fig:DrKF_4DrA}. We concentrate here on the last measurement scenario with larger angular increments, as it delivers a good reconstruction quality under a realistic imaging scenario.
We note, that the wood logs scanned were not dried before measurements were taken and therefore the external wet rings of the tree \rev{have an image intensity} indistinguishable from the knot parts. This wet part can be clearly seen in the segmentations and represents a good portion of the selected pixels. 
As a consequence, the Dice squared coefficients \cite{SDI} 

is influenced both by the wet parts and by the knots. 
Nevertheless, the segmentation method used is able to locate the knots adequately already with 3 sources and it does not improve significantly for schemes having more \rev{than} 5 sources. A similar behaviour was observed for the random angular sampling.
This is confirmed in \cref{fig:coeff_comparisons}, which further shows that the last acquisition scheme presented has a comparable precision to the random rotation case, for all the different number of sources considered. 
Where only a slight disadvantage is observed for less than 5 angles.

\begin{figure}[!ht]
\centering
\subfloat[Reference A]{\includegraphics[width=0.32\columnwidth]{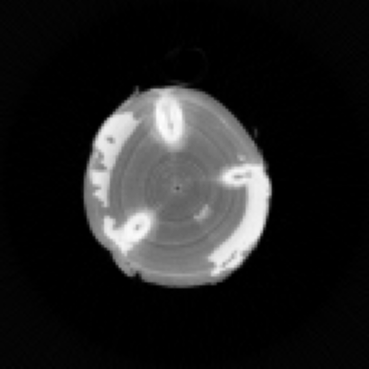}}
\hspace{0.1cm}
\subfloat[DPA segmentation ]{\includegraphics[width=0.32\columnwidth]{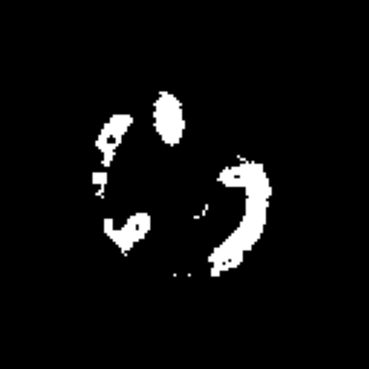}}
\hspace{0.1cm}
\subfloat[multi OTSU segmentation ]{\includegraphics[width=0.32\columnwidth]{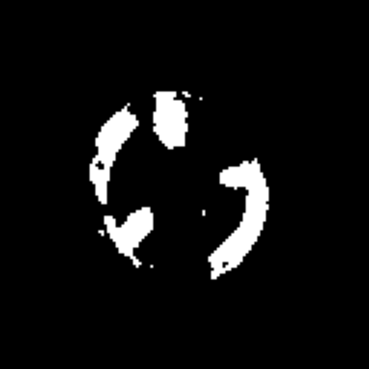}}
\caption{\rev{Segmentations of the Reference A by using DPA and a standard multi OTSU. }  }
\label{fig:fig_comp_seg}
\end{figure}

\rev{We compared the segmentation performance obtained by using DPA with a standard method such as multi-OTSU \cite{Liao2001AFA}. Despite the performance of the two methods is statistically comparable \cref{fig:comparisons}, DPA performs better on the fine details where multi OTSU tends to overestimate the \rev{knot} 
, \cref{fig:fig_comp_seg}.}

\section{Discussion}\label{sec:discussion}

In the previous section, we have shown how \rev{the} rotation of the acquisition geometries affects greatly the reconstruction quality. We studied four different schemes including rotating and non rotating acquisition geometry. In the non rotating case, we saw that there is no gain in using the DrKF formulas. We saw also that the speed of rotation needs to be chosen properly in order to get the best out of this approach. The take home message is that to exploit the potential of the filtering formulas we should cover, with the number of sources available, the angular sampling of the measurements for each 2D section as efficiently as possible within approximately 5 slices (about 1 cm in the vertical direction). 
The results given in \cref{fig:coeff_comparisons} \rev{indicate} that, if the rotation speed is tuned properly, one could obtain fair results by using as little as 3 X-ray source-detector pairs and we do not get significant gains by using more than 7. 
Alternatively, to the fixed rotation speed, one could also consider \rev{finding} an optimal sampling strategy for the scanning procedure \cite{burger2021sequentially}, but which would add a significant computational overhead to the reconstruction.

\subsection{Limitations}

\begin{figure}[!ht]
    \centering
        \subfloat[Comparison DPA - multi OTSU. ] {\includegraphics[width=0.45\columnwidth]{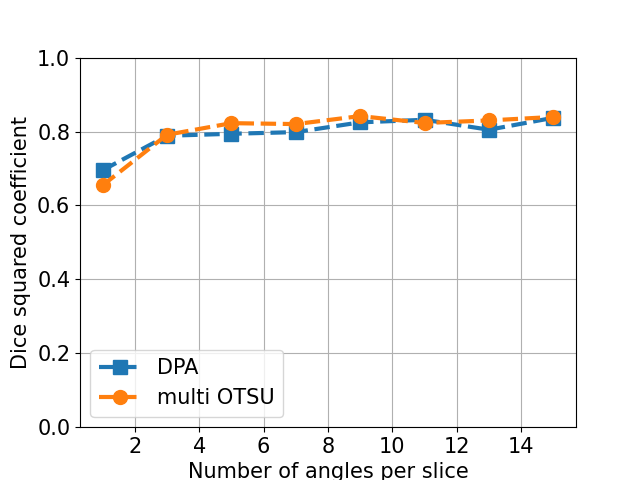}}
        \hspace{0.1cm}
        \subfloat[Impact of the number $r$ of singular values used in the dimension reduction.]{\includegraphics[width=0.45\columnwidth]{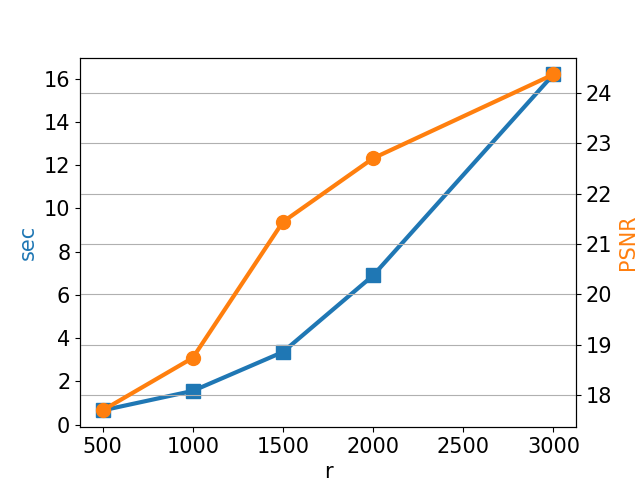}}

    \caption{\rev{Values are obtained by averaging in the block of 11 slices centred at the reference in fig. \ref{fig:DrKF_REFERENCES_A}. (A) Comparison of the segmentation method used with a standard multi OTSU. (B) Impact of the reduced dimension $r$ (amount of singular values) on the computational (blue) time and the reconstruction quality (orange).}}
    \label{fig:comparisons}
\end{figure}

Despite the good performance obtained in terms of precision, we must state also that the computational times per slice in the current implementation \rev{are} too high for online use.
For the experiments presented here, most of the computational effort $\sim$15 sec. is used to obtain one slice reconstructed while 0.7 sec. \rev{is on } average needed to segment a block of 10 slices.
\rev{If the dimensions to which the image is projected are cut more drastically before performing the KF steps to $r=1000$, the time per slice reconstructed could be reduced by approximately 10 times, but with major loss in reconstruction quality from 24 dB to 18 dB. We present in \cref{fig:comparisons} a summary of the impact of this choice the computational cost and on the reconstruction quality.}
\rev{The computations were performed with Python on a Linux workstation with 128GB memory and a Dual CPU (Intel(R) Xeon(R) Silver 4116).}

We note that this time values take into account that we pre-computed all the system and projection matrices needed in the DrKF formulas, as assembling each matrix representation to solve the update equations for the Kalman filter would require over a minute.

Finally, it should be noted that the need for \rev{a} correct rotation between slices is a practical limitation and needs to be addressed to allow for a high throughput system.

\section{Conclusions}\label{sec:conclusion}

This work proposes a Kalman filtering method for reconstruction and an unsupervised clustering approach to spot density anomalies in the internal structure of objects. The internal 3D structure of the object is obtained  by 
accumulating information from severely sparse angular sequential sampling in 2D, done as the object passed \rev{through} the X-ray scanner.
We have shown how the rotation of the measurement device plays a key role in the reconstruction quality. 
In particular, we presented a realistic constant rotation speed scheme that maintained the same level of accuracy as the best theoretical scenario, namely a random angular incremental speed, which would be impractical for practical use. 

Furthermore, we saw that there is no need to use more \rev{than} 7 sources if the acquisition scheme is designed properly and that one could achieve good results by using as little as 3 X-ray sources and a fairly simple measurement device. 
The unsupervised clustering approach used \rev{was revealed} to be robust against reconstruction noise.
As the logs were not dried before the scanning process, the Dice coefficients presented in \cref{fig:coeff_comparisons} are significantly affected by the wet external part of the tree that \rev{gets} mixed with the wood knots given that they have very similar \rev{image intensity}.
This could be overcome by spectral data due to different energy \rev{attenuation} coefficients and left as a possibility to investigate in future work.

Finally, an option to overcome the computational burden could be by extending the Kalman filtering framework in combination with learning based methods. For instance by replacing \rev{computationally} expensive parts with a neural network or intertwining model and data-driven components in the algorithm \cite{arjas2022neural,revach2022kalmannet}. Nevertheless, the nature of accumulating information between slices by a well chosen angular sampling is the essential part and needs to be carried over to the learning-based setting.

\section*{Acknowledgments} 
We thank Alessandro Laio for support and helpful discussions on the Segmentation by Density Peaks Advanced used within this study. Codes for the Kalman filtered reconstructions will be published after acceptance. 


\bibliographystyle{siam}
\bibliography{references.bib}

\medskip
Received xxxx 20xx; revised xxxx 20xx; early access xxxx 20xx.
\medskip

\end{document}